\documentclass[11pt]{article}
\usepackage{arabtex}
\usepackage{coling2018}
\usepackage{utf8}
\setcode{utf8}
\usepackage{url}
\usepackage{graphicx} 

\usepackage[stable]{footmisc}
\date{}

\begin{document}
	
	\title{\textbf{Comparing the Accuracy of Deep Neural Networks (DNN) and Convolutional Neural Network (CNN) in Music Genre Recognition (MGR): Experiments on Kurdish Music}}

	\author{
		\begin{tabular}[t]{c}
			Aza Zuhair \and Hossein Hassani\\
			\textnormal{University of Kurdistan Hewl\^er}\\
			\textnormal{Kurdistan Region - Iraq}\\
			{\tt {\{aza.zuhairamin, hosseinh\}@ukh.edu.krd}}
		\end{tabular}
	}
	
	\maketitle
	
\begin{abstract}
	Musicologists use various labels to classify similar music styles under a shared title. But, non-specialists may categorize music differently. That could be through finding patterns in harmony, instruments, and form of the music. People usually identify a music genre solely by listening, but now computers and Artificial Intelligence (AI) can automate this process. The work on applying AI in the classification of types of music has been growing recently, but there is no evidence of such research on the Kurdish music genres. In this research, we developed a dataset that contains 880 samples from eight different Kurdish music genres. We evaluated two machine learning approaches, a Deep Neural Network (DNN) and a Convolutional Neural Network (CNN), to recognize the genres. The results showed that the CNN model outperformed the DNN by achieving 92\% versus 90\% accuracy.
	
\end{abstract}
\section{Introduction}
Music is classified into genres that share the same style, melody, and culture.
Genre is a conceptual tool to classify music and other forms of art \cite{lena2008classification}. According to \newcite{norowi2005factors}, automated music genre classification or music genre recognition (MGR) is to structure and organize a very large music archive using computers. MGR is used in various applications such as Music Information Retrieval (MIR) \cite{downie2003music} and recommendation systems \cite{lorince2016consumption}, for example, Tidal, Spotify, and Apple Music. 

In the absence of automated approaches, musicologists classify pieces of music into different genres based on lyrics and melody just by listening to them. However, with the growing collection of music genres and music itself, the need for automated music classification has also increased \cite{pupp2021}. Consequently, studying the usage of advanced machine learning techniques such as neural networks in the development of more efficient and accurate music genre classification has gained more attention in the computer science domain.  

In this research, we apply and evaluate the accuracy of two neural network approaches, Deep Neural Network (DNN) and a Convolutional Neural Network (CNN), in automatically classifying Kurdish music into its different genres. The rest of this paper is organized as follows. Section \ref{sec:rw} reviews the related work. Section \ref{sec:method} provides the method of data collection, training, and evaluation. In Section \ref{sec:res}, we present and discuss the outcomes of the experiments. Finally, Section \ref{sec:con} concludes the paper.

\section{Related Work}
\label{sec:rw}

\newcite{ghosal2018music} compared two different machine learning models, CNN and a Long Short-Term Memory (LSTM), and for music classification. They compared the performance of the models on different types of features such as Mel-Spectrogram, Mel Coefficients, and Tonnetz Features.

\newcite{feng2014deep} used Deep Belief Network (DB), an unsupervised machine learning, to recognize two to four music genres. They used the GTZAN dataset that consisted of 1000 pieces of music from 10 different genres to train the model. By extracting the Mel Frequency Cepstral Coefficient (MFCC) from the music, they generated 15 samples per music and created a dataset of 15000 samples to train and test the model (60\% for training and 40\% for testing). Their DBF model consisted of five layers. The related layers were trained over Restricted Boltzmann Machine (RBM) iteratively. The model achieved 98.15\% accuracy in recognizing two genres, 69.16\% in recognizing three genres, and 51.88\% in recognizing four genres.

\newcite{silla2008machine} used an ensemble learning method for MGR based on space and time decomposition and evaluated the results of different segment parts. They divided each music into three different segments: the music's beginning, middle, and end. Ensemble learning is a combination of different learning algorithms to enhance the classification result. The learning algorithms in the mentioned experiment were Decision Tree, K-Nearest Neighbours (KNN), Naive-Bayes, Multi-Layer Perceptron neural network, and Support Vector Machine (SVM). The results showed the middle segment of the music performing better than the other two segments.

\newcite{tzanetakis2002musical} applied Gaussian Mixture Model and KNN for MGR. They used three features of the audio, \textit{pitch}, \textit{rhythm}, and \textit{timbral texture} in their model.

\newcite{scheirer1997construction} developed a multidimensional classifier based on several models, such as Maximum a Posteriori (MAP), Gaussian Mixture Model (GMM), and KNN in their MGR experiments in which they incorporated 13 different features of the music. They trained the classifier on a frame-by-frame audio segment basis that achieved an error rate of 5.8\%. However, by extending the length of the segments the error rate dropped down to 1.4\%.

To summarize, the literature shows using a wide range of machine learning methods in MGR. It also shows the variety of methods in data segmentation and datasets preparation. We also observe growing attention to MGR in general, but we did not notice any study on Kurdish MGR during the literature review and while compiling this paper.   

\section{Methodology}
\label{sec:method}

\subsection{Data Collection}
We reach out to fine art teachers, students, and artists, to explore various Kurdish music genres and understand them. Because there is no ready-to-use dataset, we collect music from various sources online and offline, such as Youtube, Soundcloud, and Music CDs. We organize the collected music into various genres. \newcite{ahmadi2020corpus} mentioned some of the Kurdish music genres, for example, Bend, Gorani Meqam, and Hayran. We collect the data in Waveform Audio File Format (WAVE or WAV) where they are available, otherwise in MPEG-1 Audio Player III (MP3) format. However, we convert the MP3s into WAV to have a decompressed version of the music.

\subsection{Feature Extraction}
For feature extraction, we used Librosa version (0.8.1) \cite{mcfee2015librosa}, a python library for analyzing music and audio that is often used in the field of MRI. We use MFCC\footnote{\url{https://musicinformationretrieval.com/mfcc.html}} features as the sole attribute for classifying the Kurdish genres. MFCC is for describing the timbre of the music, and it consists of a small set of features that describe the shape of the spectral envelope. Using Librosa, we segment the 30-second audio files into 1200 segments, which results in the audio files being much smaller but the dataset much bigger for training. However, according to \newcite{scheirer1997construction}, longer segments perform better at classifying audio, and therefore, the number of segments extracted from the music might change, depending on the result of the experiments.

\subsection{Data Preparation}
Initially, we randomize the dataset and split it into training, validation, and testing datasets that receive 70\%, 10\%, and 20\% of the original dataset accordingly. However, the splitting ratio might change depending on the length of the collected dataset and the results that we obtain.

\subsection{Architecture of the Models}
We use two architectures using TensorFlow\footnote{\url{https://www.tensorflow.org/}} libraryand Keras\footnote{\url{https://keras.io/}} API. The architectures are based on those that \newcite{ref16} suggests. We may modify the architecture and the number of epochs during the experiments to improve the performance of the models.

\begin{figure}[hbt!]
	\begin{center}
	\includegraphics[width=0.6\textwidth]{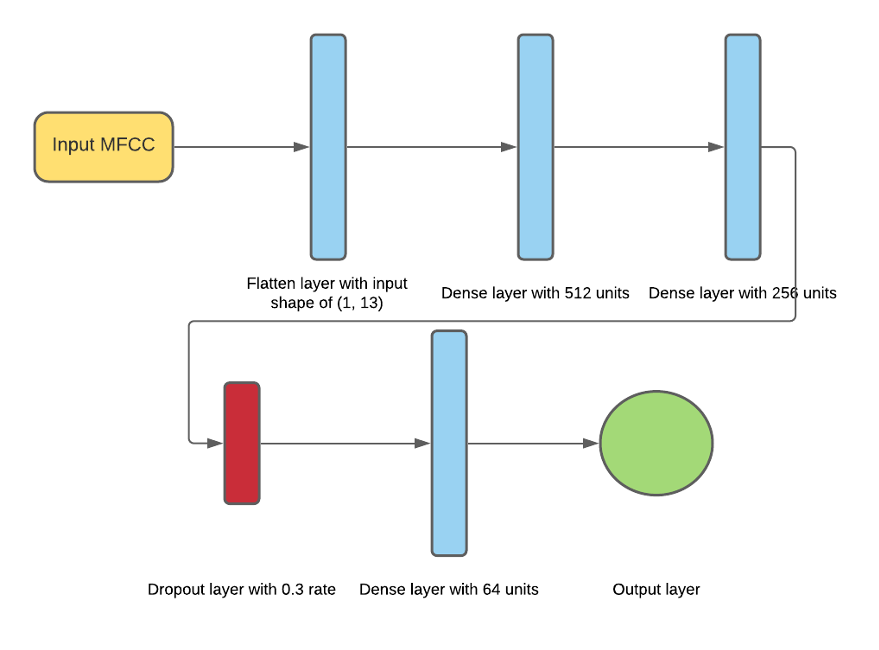}
	\caption{Modified Deep Neural Network architecture.} \label{fig4}
	\end{center}
\end{figure}

The first model shown in Figure~\ref{fig4} is a Deep Neural Network that consists of four layers. For the input layer, we flatten the data with the shape of (1, 13), and then we fed it into the first layer that has 512 units. We use Rectified Linear (ReLu) as the activation function. The second layer has 256 units, and it uses the same activation function as the first layer. Between the second layer and the third layer, we insert a dropout layer with a rate of 0.3 to prevent overfitting in the model. The third layer consists of 64 units, and it uses the same activation function as the first and second layers. Finally, the output layer uses a \textit{softmax} activation function to classify the given data. 

\begin{figure}[hbt!]
	\begin{center}
	\includegraphics[width=0.6\textwidth]{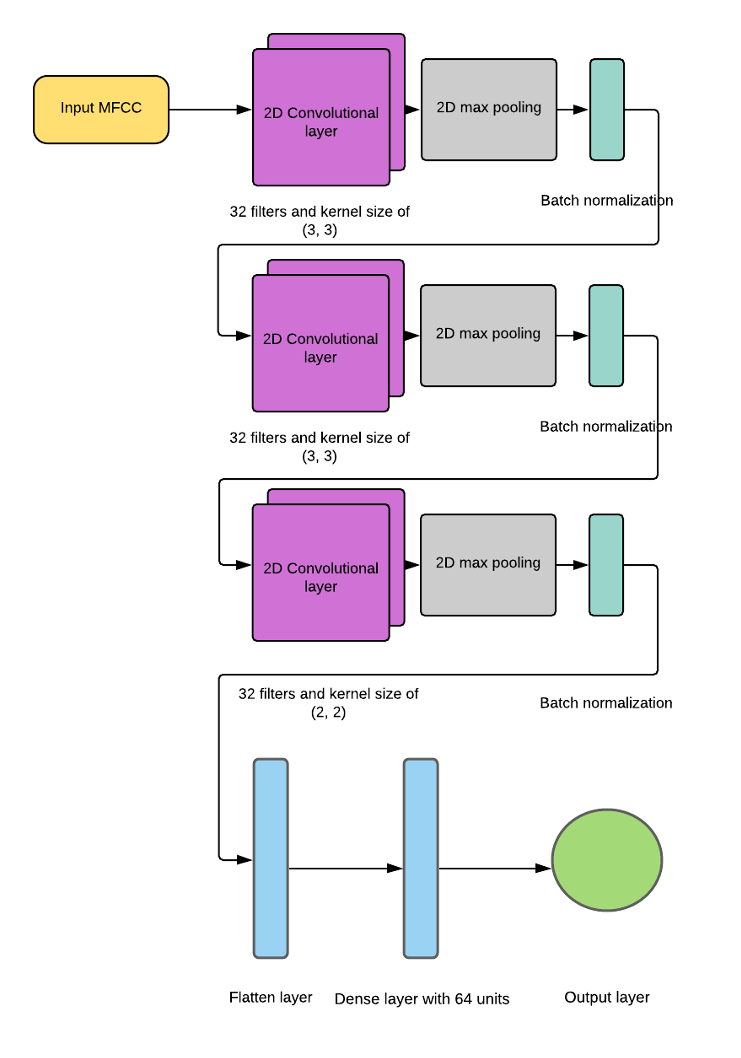}
	\caption{Modified Convolutional Neural Network architecture.} \label{fig5}
\end{center}
\end{figure}

Figure~\ref{fig5} shows the second model, CNN, which consists of five layers. The first layer is a two-dimensional (2D) convolutional layer with 32 filters, kernel size of (3, 3), and it uses ReLu as the activation function. The model also contains a 2D Max Pooling layer with a pool size of (3, 3) and strides of (2, 2). We fed the data into a batch normalization layer to normalize the inputs. The second layer is the same as the first layer, consisting of 2D convolutional, 2D max pooling, and batch normalization layers. The third layer contains a 2D convolutional layer with 32 filters, kernel size of (2, 2), and uses ReLu as the activation function. A 2D max pooling layer with a pool size of (2, 2) comes after the 2D convolutional layer. We fed the output of the max-pooling layer into a batch normalization layer. The fourth layer flattens the output of the last layer and feeds it into a dense layer. The dense layer consists of 64 units and uses ReLu as the activation function. The last layer uses a \textit{softmax} function to predict the genre. 

\subsection{Evaluation and Testing}
We evaluate the accuracy of the models using a TensorFlow metric that is called Accuracy\footnote{\url{https://www.tensorflow.org/api_docs/python/tf/keras/metrics/Accuracy}}. Formula \ref{formula-1} shows the metric, in which \textit{accuracy} is the calculated accuracy of the model, \textit{frequency} is the frequency of the predictions that match the \textit{true} label, and \textit{total samples} is the number samples examined in the experiment.

\begin{equation}
\label{formula-1}
accuracy = frequency / total samples
\end{equation}

\section{Experiments and Result}
\label{sec:res}

\subsection{Data Collection}
We collected 208 pieces of music from eight genres. The length of the pieces is between one to 30 minutes. We trimmed the music into 30-second audio files. We followed the method suggested by \newcite{silla2008machine}, and we took the samples from the middle to keep the overall timbre of the music. The trimming process provided 2293 samples. However, the data was still imbalanced because \textit{Gorani Classic} genre had a larger portion of data in comparison to the other genres. That is why only an equal amount of music in each genre had to be selected for training and testing the models. Therefore, we randomly selected 110 pieces because the minimum number of samples that we had was 110 samples for one of the genres \textit{Halparke Se-pey} genre that had 110 pieces of music after the trimming process.

\begin{figure}[hbt!]
	\begin{center}
	\includegraphics[width=0.75\textwidth]{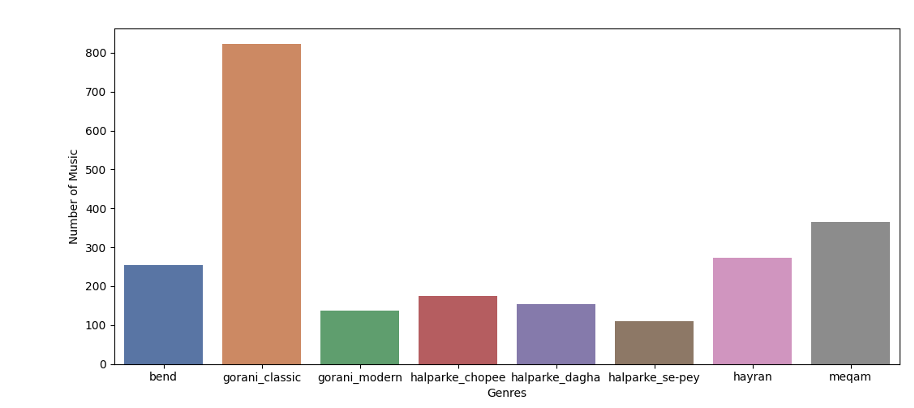}
	\caption{Bar chart representing the number of music in each genre after pre-processing.} \label{fig1}
	\end{center}
\end{figure}

During our experimentation, we observed the results and made modifications to improve the models. Tables~ \ref{tab1}, \ref{tab2}, and \ref{tab3} show those modifications along with the attributes for training and evaluating the models for each experiment. Table~\ref{tab1} summarizes the experiments we did on the DNN model, with the results of each experiment.

\begin{table}[ht!]
	\begin{center}
	\caption{Experiments and results of the DNN model.}\label{tab1}
	\begin{tabular}{|l|l|l|l|l|l|l|}
	\hline
	No. &  Input Shape & Total Samples & Randomization Algorithm & Epochs &  Accuracy & Loss\\
	\hline
	1 &  (2, 13) & 1,056,000 & Random Split & 30 & 74\% & 72\%\\
	2 &  (2, 13) & 200,000 & Random Split & 30 & 67\% & 90\%\\
	3 &  (2, 13) & 1,056,000 & k-Fold Cross-validation & 30 & 74\% & 71\%\\
	4 &  (2, 13) & 1,056,000 & k-Fold Cross-validation & 100 & 75\% & 69\%\\
	5 &  (44, 13) & 26,400 & k-Fold Cross-validation & 30 & 90\% & 39\%\\
	6 &  (44, 13) & 26,400 & k-Fold Cross-validation & 30 & 90\% & 32\%\\
	\hline
	\end{tabular}
	\end{center}
\end{table}

Table~\ref{tab2} is the summary of the experiments and results on the CNN model.

\begin{table}[ht!]
\begin{center}
	\caption{Experiments and results of the CNN model.}\label{tab2}
	\begin{tabular}{|l|l|l|l|l|l|l|}
	\hline
	No. &  Input Shape & Total Samples & Randomization Algorithm & Epochs &  Accuracy & Loss\\
	\hline
	1 &  (2, 13, 1) & 1,056,000 & Random Split & 30 & 67\% & 88\%\\
	2 &  (2, 13, 1) & 200,000 & Random Split & 30 & 64\% & 98\%\\
	3 &  (2, 13, 1) & 1,056,000 & k-Fold Cross-validation & 30 & 67\% & 88\%\\
	4 &  (2, 13, 1) & 1,056,000 & k-Fold Cross-validation & 100 & 69\% & 85\%\\
	5 &  (44, 13, 1) & 26,400 & k-Fold Cross-validation & 30 & 92\% & 23\%\\
	6 &  (44, 13, 1) & 26,400 & k-Fold Cross-validation & 30 & 92\% & 25\%\\
	\hline
	\end{tabular}
\end{center}
\end{table}

Table~\ref{tab3} is the details of the datasets that are used for each experiment on both models. As the results show, we achieved 90\% accuracy for the DNN model and 92\% accuracy for the CNN model. 

\begin{table}[ht!]
	\begin{center}
	\caption{Details of the dataset for each experiments.}\label{tab3}
	\begin{tabular}{|l|l|l|l|l|l|l|}
	\hline
	No. & Total Samples & Splitting Ratio & Sample Length &  Train Set & Valid Set & Test Set\\
	\hline
	1 & 1,056,000 & 7:1:2 & 25ms & 739,200 & 105,600 & 211,200\\
	2 & 200,000 & 5:2.5:2.5 & 25ms & 100,000 & 50,000 & 50,000\\
	3 & 1,056,000 & 7:1:2 & 25ms & 739,200 & 105,600 & 211,200\\
	4 & 1,056,000 & 7:1:2 & 25ms & 739,200 & 105,600 & 211,200\\
	5 & 26,400 & 8:1:1 & 1s & 21,120 & 2,640 & 2,640\\
	6 & 26,400 & 8:1:1 & 1s & 21,120 & 2,640 & 2,640\\
	\hline
	\end{tabular}
	\end{center}
\end{table}

\begin{figure}[hbt!]
	\begin{center}
		\includegraphics[width=0.72\textwidth]{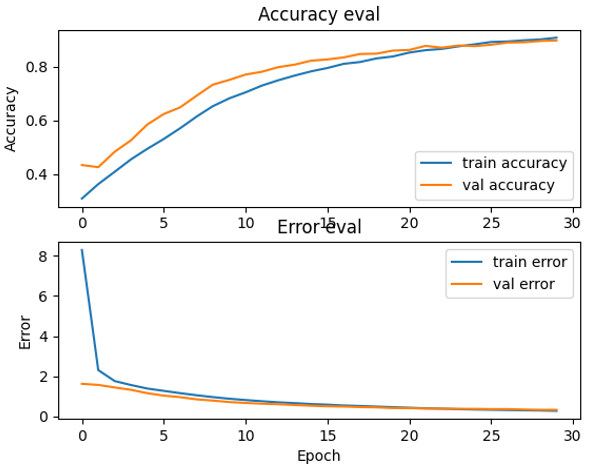}
		\caption{Sixth experiment training process for the DNN model.} \label{fig6}
	\end{center}
\end{figure}

\begin{figure}[hbt!]
	\begin{center}
		\includegraphics[width=0.8\textwidth]{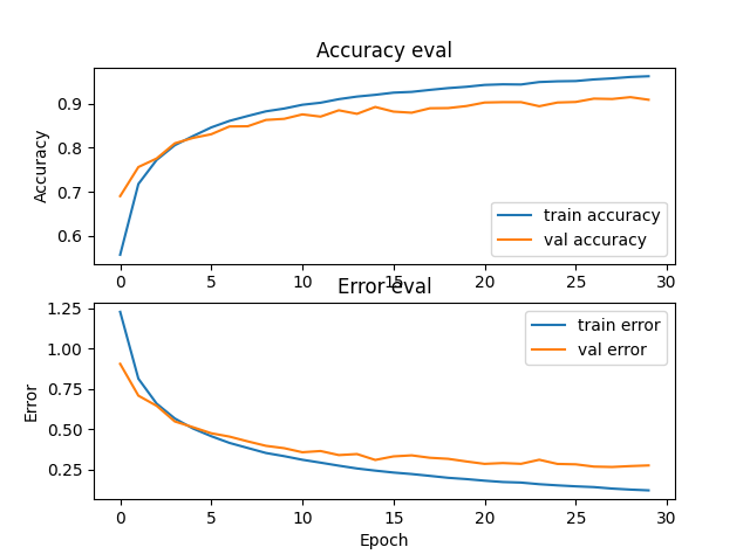}
		\caption{Fifth experiment training process for the CNN model.} \label{fig7}
	\end{center}
\end{figure}

Fig.~\ref{fig6} shows the training performance of the DNN model in the sixth experiment, and Figure Fig.~\ref{fig7} shows it for the CNN model in the fifth experiment.

\subsection{Discussion}

Although we achieved 90\% accuracy for DNN and 92\% for CNN, when we look back at the first four experiments of the models, we can see that they performed poorly on the testing dataset. For both models, the first experiment did not perform well on the testing dataset, and the loss was more than 70\% in both cases. 

To investigate the problem, in the second experiment, we trained and tested the models on a smaller dataset, but the results did not improve; the accuracy was decreased, and the loss was increased. While training the models in the first and second experiments, the validation accuracy was higher than the training accuracy, indicating that the dataset is not well-randomized.

In the third experiment, we changed the algorithm for randomizing the dataset to the cross-validation algorithm mentioned in the methodology. In this experiment, the dataset was distributed evenly for each genre using the StratifiedKFold class in the Scikit-learn library\footnote{\url{https://scikit-learn.org/stable/}}. We used the last distributed group (Fold) as the dataset for conducting the rest of the experiments. This time, the results showed a higher validation accuracy than the training accuracy similar to the second experiment. 

\newcite{ref18} states that if the model’s accuracy does not decrease across the epochs, increasing the number of epochs could increase the accuracy. For the fourth experiment, we increased the number of epochs to 100 for both models. The DNN model validation accuracy was less than the training, but it did not perform well on detecting the genres. Unlike the DNN, the CNN model result was similar to the previous experiment, the validation accuracy being higher than the training accuracy.

Also, according to \newcite{scheirer1997construction}, the classifier has less error rate when working with longer samples, for example, samples of 2.4 seconds length. To extract longer samples, we needed to decrease the amount of the segmentation. Therefore, we went back to feature extraction in the fifth experiment and changed the number of samples to 30 from each 30-second music. Segmenting each 30-second music into 30 samples made each sample 1 second long. We did not extract samples longer than 1 second because the dataset would be much smaller for experimenting. Additionally, because the length of the dataset would have become less than the previous experiments, the model could overfit. Thus, we changed the number of epochs back to 30.

The fifth experiment showed a promising result, with 90\% accuracy for DNN and 92\% for CNN. However, The training accuracy was higher than the validation accuracy by seven percent, indicating a slight overfit in the model. To fix the minor overfit, we made a slight modification to the architecture of the models, adding one dropout layer with the rate of 0.3 to both models. After that, both of the models showed no overfit.

Looking back at the experiments, the best result for the DNN model was in the sixth experiment. We achieved 90\% accuracy with a 32\% loss. Additionally, the best result for the CNN model was in the fifth experiment because it had less error rate than the sixth error rate, which means it made fewer errors on the testing dataset. The model was 92\% accurate with a 23\% loss. 

\section{Conclusion}
\label{sec:con}

As music becomes more available in large collections, their genre recognition gets more attention. As a result, Music Genre Recognition (MGR) has become a field of study. Scientists have attempted to classify music genres using machine learning and Artificial Intelligence (AI). However, so far, using those techniques and methods has not been practiced. In this research, we applied machine learning techniques to recognize Kurdish music genres by using two artificial neural network methods, DNN and CNN,  using machine learning. We collected and built a dataset of 880 music from eight different Kurdish music genres. Each genre has 110 music, and each piece of music is 30 seconds long. We developed the models and evaluated them. The CNN model scored 92\% accuracy, and the DNN model achieved 90\% accuracy, which states that the CNN model outperformed the DNN model at recognizing the Kurdish music genres.

The Kurdish MGR requires extensive efforts in the future. Some of the areas could be of immediate interest., for example, to train the models on better quality music. Some of the music that we used in this research were not of high-quality sound. Also, working on the recognition of low-quality music could be a research area because one could find a considerable amount of Kurdish music that has been recorded using low-quality instruments. To separate the songs from the music for each genre could be another working area that might positively affect genre recognition. Furthermore, training the models with longer samples could also be investigated. Finally, the classified dataset used for this research can be used for building a recommender system for Kurdish music genres.

%
%
%

\bibliographystyle{lrec}
\bibliography{KurdishMGR}

\begin{thebibliography}{}

\bibitem[\protect\citename{Ahmadi \bgroup et al.\egroup
  }2020]{ahmadi2020corpus}
Ahmadi, S., Hassani, H., and Abedi, K.
\newblock (2020).
\newblock A corpus of the sorani kurdish folkloric lyrics.
\newblock In {\em Proceedings of the 1st Joint Workshop on Spoken Language
  Technologies for Under-resourced languages (SLTU) and Collaboration and
  Computing for Under-Resourced Languages (CCURL)}, pages 330--335.

\bibitem[\protect\citename{Downie}2003]{downie2003music}
Downie, J.~S.
\newblock (2003).
\newblock Music information retrieval.
\newblock {\em Annual review of information science and technology},
  37(1):295--340.

\bibitem[\protect\citename{Feng}2014]{feng2014deep}
Feng, T.
\newblock (2014).
\newblock Deep learning for music genre classification.
\newblock {\em private document}.

\bibitem[\protect\citename{Ghosal and Kolekar}2018]{ghosal2018music}
Ghosal, D. and Kolekar, M.~H.
\newblock (2018).
\newblock Music genre recognition using deep neural networks and transfer
  learning.
\newblock In {\em Interspeech}, pages 2087--2091.

\bibitem[\protect\citename{Lena and Peterson}2008]{lena2008classification}
Lena, J.~C. and Peterson, R.~A.
\newblock (2008).
\newblock Classification as culture: Types and trajectories of music genres.
\newblock {\em American sociological review}, 73(5):697--718.

\bibitem[\protect\citename{Lorince}2016]{lorince2016consumption}
Lorince, J.
\newblock (2016).
\newblock {\em Consumption of content on the web: An ecologically inspired
  perspective}.
\newblock {Ph.D.} thesis, Indiana University.

\bibitem[\protect\citename{McFee \bgroup et al.\egroup }2015]{mcfee2015librosa}
McFee, B., Raffel, C., Liang, D., Ellis, D.~P., McVicar, M., Battenberg, E.,
  and Nieto, O.
\newblock (2015).
\newblock librosa: Audio and music signal analysis in python.
\newblock In {\em Proceedings of the 14th python in science conference},
  volume~8, pages 18--25. Citeseer.

\bibitem[\protect\citename{Norowi \bgroup et al.\egroup
  }2005]{norowi2005factors}
Norowi, N.~M., Doraisamy, S., and Wirza, R.
\newblock (2005).
\newblock Factors affecting automatic genre classification: an investigation
  incorporating non-western musical forms.
\newblock In {\em Proceedings of the International Conference on Music
  Information Retrieval}, pages 13--20.

\bibitem[\protect\citename{Puppala \bgroup et al.\egroup }2021]{pupp2021}
Puppala, L.~K., Muvva, S. S.~R., Chinige, S.~R., and Rajendran, P.
\newblock (2021).
\newblock A novel music genre classification using convolutional neural
  network.
\newblock In {\em 2021 6th International Conference on Communication and
  Electronics Systems (ICCES)}, pages 1246--1249.

\bibitem[\protect\citename{Scheirer and Slaney}1997]{scheirer1997construction}
Scheirer, E. and Slaney, M.
\newblock (1997).
\newblock Construction and evaluation of a robust multifeature speech/music
  discriminator.
\newblock In {\em 1997 IEEE international conference on acoustics, speech, and
  signal processing}, volume~2, pages 1331--1334. IEEE.

\bibitem[\protect\citename{Silla \bgroup et al.\egroup }2008]{silla2008machine}
Silla, C.~N., Koerich, A.~L., and Kaestner, C.~A.
\newblock (2008).
\newblock A machine learning approach to automatic music genre classification.
\newblock {\em Journal of the Brazilian Computer Society}, 14(3):7--18.

\bibitem[\protect\citename{Taub}2021]{ref18}
Taub, D.
\newblock (2021).
\newblock {Must accuracy increase after every epoch?}
\newblock Available at \url{https://stackoverflow.com/questions/45605003/must-
  accuracy-increase-after-every-epoch}.

\bibitem[\protect\citename{Tzanetakis and Cook}2002]{tzanetakis2002musical}
Tzanetakis, G. and Cook, P.
\newblock (2002).
\newblock Musical genre classification of audio signals.
\newblock {\em IEEE Transactions on speech and audio processing},
  10(5):293--302.

\bibitem[\protect\citename{Velardo}2021]{ref16}
Velardo, V.
\newblock (2021).
\newblock {DeepLearningForAudioWithPython}.
\newblock Available at
  \url{https://github.com/musikalkemist/DeepLearningForAudioWithPython}.

\end{thebibliography}

\end{document}